\documentclass[aps,prl,twocolumn,showpacs,superscriptaddress,nofootinbib,groupedaddress]{revtex4} 
\usepackage{graphicx}
 \usepackage{epsfig}
\usepackage{tabularx}
\usepackage{booktabs}
\usepackage{amssymb}
\usepackage{amsmath}
\RequirePackage[colorlinks,citecolor=blue,urlcolor=blue,linkcolor=blue]{hyperref}
\usepackage{dcolumn}
\begin{document}

\preprint{PRD}

\title{The Cabibbo angle as a universal seed for quark and lepton mixings}

\author{S.~Roy}
 \email{meetsubhankar@gmail.com}
\affiliation{Department of Physics, Gauhati University, Guwahati, Assam -781014, India
}%
\author{S.~Morisi}
 \homepage{stefano.morisi@gmail.com}
\affiliation{
DESY, Platanenallee 6, D-15735 Zeuthen, Germany
}%
\author{N.~N.~Singh}
 \email{nimai03@yahoo.com}
\affiliation{Department of Physics, Manipur University, Imphal, Manipur-795003, India
}%
\author{J.~W.~F.~Valle}
 \homepage{valle@ific.uv.es}
\affiliation{
AHEP Group, Institut de F\'{i}sica Corpuscular --
  C.S.I.C./Universitat de Val\`{e}ncia, Parc Cientific de Paterna.\\
 C/ Catedratico Jos\'e Beltr\'an, 2 E-46980 Paterna (Val\`{e}ncia) - SPAIN
}%
\date{\today}


 
\begin{abstract}

  A model-independent ansatz to describe lepton and quark mixing in a
  unified way is suggested based upon the Cabibbo angle. In our
  framework neutrinos mix in a ``Bi-Large'' fashion, while the charged
  leptons mix as the ``down-type'' quarks do. In addition to the
  standard Wolfenstein parameters ($\lambda$, $A$) two other free
  parameters ($\psi$, $\delta$) are needed to specify the physical
  lepton mixing matrix. Through this simple assumption one makes
  specific predictions for the atmospheric angle as well as 
  leptonic CP violation in good agreement with current
  observations.

\end{abstract}

\maketitle


A striking observation vindicated by recent experimental neutrino data
is that the smallest of the lepton mixing angles is surprisingly
large, similar to the largest of the quark mixing parameters, namely
the Cabibbo angle ($\theta_c$)~\cite{Tortola:2012te,Forero:2014bxa}.
An interesting lepton mixing scheme called ``Bi-Large'' (BL) mixing
has been proposed recently~\cite{Boucenna:2012xb} and subsequently
studied in Refs.~\cite{Ding:2012wh,Branco:2014zza,Roy:2012ib}.  This
mixing scheme assumes the atmospheric and the solar mixing angles to
be equal and proportional to the reactor angle.  
In contrast to the Bi-maximal (BM)
scenario~\cite{barger:1998ta,altarelli:1998nx}, within the BL scheme
the atmospheric mixing angle does not need to be strictly
``Maximal'', but simply ``Large'' in general.  In summary, BL mixing
posits, $\sin\theta_{13}\simeq\lambda$,
$\sin\theta_{12}=\sin\theta_{23}\sim\lambda$, where
$\lambda=\sin\theta_{c}$.

Such BL mixing ansatz can be motivated in string theories. Indeed, in
F-theory motivated Grand Unified Theory (GUT) models, a geometrical
unification of charged lepton and neutrino sectors leads to a mild
hierarchy in the neutrino mixing matrix in which $\theta_{12}^{\nu}$
and $\theta_{23}^{\nu}$ become large and comparable while
$\theta_{13}^{\nu}\sim\theta_{c}\sim \sqrt{\alpha_{GUT}}\sim 0.2$
\cite{bouchard2010f}~\footnote{Neglecting the contribution from the
  charged lepton sector.}.
Understanding the origin of the above relation from first principles
is beyond the scope of this note. We stress however that this ansatz
can be associated to specific flavor symmetries as suggested in
Ref.~\cite{Ding:2012wh} or Ref.~\cite{King:2012in}, rather than being
a mere ``numerical coincidence''.

A successful framework for attacking the flavour problem constitutes
an important quest in contemporary particle physics. A relevant
question arises as to whether attempted solutions to the flavour
problem may indicate foot-prints of unification or not. In the present
note we look into some possible links between quark and lepton mixing
parameters from a phenomenological ``bottom-up
perspective''\,\footnote{An earlier alternative in the literature is
  ``Quark-Lepton complementarity (QLC)''
  \cite{raidal:2004iw,minakata-2004-70,ferrandis:2004vp,Zhang:2012xu}. }.\\

In the quark sector the largest mixing is between the flavor states
$d$ and $s$, and is interpreted in terms of the Cabibbo
angle~\cite{cabibbo:1978nk} which is approximately $13^0$.  The matrix
$V_{CKM}$ is parametrized in terms of three independent angles and one
complex CP
phase~\cite{kobayashi:1973fv,Schechter:1980gr,Rodejohann:2011vc}. A
clever approximate presentation was proposed by
Wolfenstein~\cite{PhysRevLett.51.1945}, and is by now standard, namely
\begin{eqnarray}
V_{CKM} = \begin{bmatrix}
1-\frac{1}{2}\lambda^2 & \lambda & A\lambda^3(\rho-i\eta) \\ 
-\lambda & 1-\frac{1}{2}\lambda^2 & A\lambda^2 \\ 
A\lambda^3(1-\rho-i\eta) & -A\lambda^2 & 1
\end{bmatrix}  
\end{eqnarray} 
up to $\mathcal{O}(\lambda ^4)$ where, $\lambda$, $A$, $\eta$ and
$\rho$ are four independent Wolfenstein parameters, with
$\lambda=\sin\theta_{c}\approx 0.22$.\\

In contrast, the mixing in the lepton sector is very different from
quark mixing.  While the solar and atmospheric angle are large:
$\theta_{12}\approx 35^0$ and $\theta_{23}\approx 49^0 $, the $1$-$3$
mixing parameter in the lepton sector is the smallest and was believed
to vanish according to the earlier results. However in last few years
it has been established \cite{Abe:2011fz,An:2012eh,Ahn:2012nd} that
this mixing, now precisely measured, is almost as large as the $d$-$s$
mixing in quark sector, $\theta_{13}\approx 9^0 \sim
\mathcal{O}(\theta_{c})$.  This excludes the simplest proposed schemes
of neutrino mixing, which need to be revised in order to be consistent
with observation~\cite{Morisi:2013qna}.
Up to Majorana phases the Bi-Large mixing factor may be
parametrized as follows
\begin{align}
U_{BL}\approx \left[
\begin{array}{ccc}
 c(1-\frac{\lambda ^2 }{2}) & \psi\lambda(1  -\frac{\lambda ^2}{2}) & \lambda  \\
 -c\psi\lambda(1+\lambda)& c^2-\lambda ^3 \psi ^2 & \psi\lambda(1  -\frac{\lambda ^2}{2}) \\
 \lambda ^2 \psi ^2-\lambda  c^2 & -c\psi\lambda(1+\lambda) & c(1-\frac{\lambda ^2 }{2}) \\
\end{array}
\right]
\end{align}
One sees that $\sin\theta_{12}=\sin\theta_{23} \approx \psi
  \lambda$, with $\sin\theta_{13}=\lambda$. With this parametrization
  it is evident that the Cabibbo angle is the {\it seed} for the
  mixing in both the quark and the lepton sector. Here, $c\approx
  \cos\sin^{-1}(\psi\lambda)$.  In what follows we discuss the
possible forms of the charged lepton
contribution to the lepton mixing matrix.\\

As originally proposed the Bi-Large ansatz does not fit current
neutrino oscillation data, so that corrections are required.  A
possibility is that BL arises only in the flavor basis and deviations
are induced from the charged lepton sector.  Here  we consider
this case within a GUT inspired framework based upon $SO(10)$ and $SU(5)$.

In simplest SO(10) schemes the charged lepton mass matrix is
approximated to that of down type quarks, $M_{e}\sim M_{d}$. This
leads to the assumption, $U_{l}\approx V_{CKM}$. In $V_{CKM}$ the
dominant parameter is $\theta_{12}^{CKM}=\theta_{c}$, which is
followed by $\theta_{23}^{CKM}$. We classify the parametrization of
$U_{l}$ in two catagories: (i) with $1$-$2$ rotation only:
$U_{l}=U_{12}(\lambda)$ and (ii) with $2$-$3$ rotation in addition to
that of $1$-$2$, $U_{l}=U_{23}(A\,\lambda^2).U_{12}(\lambda)$. As
suggested in Ref.\,\cite{Fritzsch:1997st}, we associate a complex
phase parameter $\delta$ with $1$-$2$ rotation, so that
$U_{12}\to U_{12}(\theta_{c},\delta)$. We have,
\begin{align}
\label{type1}
U_{l_{1}}&=\Psi R^{l}_{12}(\theta_{12}^{CKM})\Psi' \nonumber \\
         &\approx\begin{bmatrix}
1-\frac{1}{2}\lambda^2 & \lambda\, e^{-i \delta} & 0 \\ 
-\lambda\, e^{i \delta} & 1-\frac{1}{2}\lambda^2 & 0 \\ 
0 & 0 & 1
\end{bmatrix} \hspace{1.3 cm} (\text{\textbf{Type-1}})\\
\nonumber\\
\label{type2}
U_{l_2}&=R^{l}_{23}(\,\theta_{23}^{CKM}\,).\Psi.R^{l}_{12}(\,\theta_{12}^{CKM}\,)\Psi' \nonumber \\
  &\approx  \begin{bmatrix}
1-\frac{1}{2}\lambda^2 & \lambda\, e^{-i \delta} & 0 \\ 
-\lambda\, e^{i \delta} & 1-\frac{1}{2}\lambda^2 & A\lambda^2 \\ 
A \lambda^3 e^{i \delta} & -A \lambda^2 & 1
\end{bmatrix}, \hspace{0.8 cm} (\text{\textbf{Type-2}})
\end{align}  
where, we have $\Psi=diag\{ e^{-i\delta/2}, e^{i\delta/2},1\}$ and
$\Psi'= \Psi^{\dagger}$.\\

Similar within simplest SU(5) scheme one expects, $M_{e}\sim
M_{d}^{T}$. This gives rise to other two possibilities which can be
expressed as in the following:
\begin{align}
\label{type3}
U_{l_3} & =\Psi R^{l^{T}}_{12}(\,\theta_{12}^{CKM}\,)\Psi' \nonumber \\
 &\approx \begin{bmatrix}
1-\frac{1}{2}\lambda^2 & -\lambda\, e^{-i \delta} & 0 \\ 
\lambda\, e^{i \delta} & 1-\frac{1}{2}\lambda^2 & 0 \\ 
0 & 0 & 1
\end{bmatrix} \hspace{1.3 cm} (\text{\textbf{Type-3}})\\
\nonumber\\
\label{type4}
U_{l_4} & = \Psi. R^{l^{T}}_{12}(\,\theta_{12}^{CKM}\,)\Psi'.R^{l^{T}}_{23}(\,\theta_{23}^{CKM}\,) \nonumber\\
&\approx \begin{bmatrix}
1-\frac{1}{2}\lambda^2 & -\lambda\, e^{-i \delta} & A\lambda^3 e^{-i\delta} \\ 
\lambda\, e^{i \delta} & 1-\frac{1}{2}\lambda^2 & -A\lambda^2 \\ 
0 & A \lambda^2 & 1
\end{bmatrix} \hspace{0.2 cm} (\text{\textbf{Type-4}})
\end{align} 

The physical lepton mixing matrix is simply
\begin{equation}
U_{lep}=U_{l}^{\dagger}. U_{BL}. I_{\phi}\,,
\end{equation}
where $U_{BL}$, represents the
Bi-Large neutrino mixing matrix and $I_{\phi}=diag(e^{i\alpha},e^{i\beta},1)$,
where $\alpha$ and $\beta $ are the two additional CP violating phases
associated to the Majorana nature of the
neutrinos~\cite{Schechter:1980gr}~\footnote{As shown in
  \cite{Schechter:1980gk} these phases are physical and affect lepton
  number violating processes such as neutrinoless double beta
  decay~\cite{wolfenstein:1981rk,valle:1982yw}. }.
In what follows we base our
discussion upon the above four different choices of the charged lepton
diagonalizing matrix choices of $U_{l}$ in
Eqs.\,(\ref{type1})-(\ref{type4}).\\

As an example here we choose the Type-4 charged lepton diagonalizing
matrix, $U_{l_4}$, (see Eq.\,(\ref{type4})) and construct the Type-4
BL based scheme,
\begin{align}
(U_{lep})_{4}&= U_{l_4}^{\dagger}. U_{BL} . I_{\phi},
\end{align}
In $(U_{lep})_{4}$, the free parameters are $\psi$ and $\delta$.
From $(U_{lep})_4$, the mixing angles are given by 
\begin{eqnarray}
\label{s213pdg}
s_{13}^{2} & \approx & \lambda ^2 \left(s_{\theta}^2+2 s_{\theta} \cos \delta + 1\right),\\
\label{s212pdg}
s_{12}^{2} &\approx & s_{\theta}^2 +\lambda ^2 \left(c_{\theta}^4+s_{\theta}^4-s_{\theta}^2\right)+2 c_{\theta}^2 \lambda s_{\theta} \cos \delta,\\
\label{s223pdg}
s_{23}^2 &\approx & s_{\theta}^2+\lambda ^2 \left(2 A c_{\theta} s_{\theta}+s_{\theta}^4+2 s_{\theta}^3 \cos \delta -s_{\theta}^2-2 s_{\theta} \cos \delta \right).\nonumber\\
&&\label{s223pdg}
\end{eqnarray}

In order to obtain the rephasing-invariant CP violation parameter
relevant for the description of neutrino oscillations we use the
relation $J_{CP}=Im[U_{e1}^{*}.U_{\mu3}^{*}.U_{\mu 1}.U_{e 3}]$ for
the Jarkslog invariant $J_{CP}$~\cite{Jarlskog:1985ht}, and obtain,
\begin{eqnarray}
\label{jcppdg}
J_{cp} \approx -c_{\theta}^2 s_{\theta}^3  \lambda \sin \delta.
\end{eqnarray}
where, $s_{\theta}=\psi\lambda$. 
It is evident that the all observables are given in terms of the
parameters, $\lambda$, $A$, $\psi$ and the unphysical phase $\delta$,
of which $\lambda$ and $A$ are the standard Wolfenstein parameters
with $\lambda\approx 0.2245$, $A=0.823$ \cite{Charles:2004jd} while
the two parameters: $\psi$ and $\delta$ are free.\\

\begin{table*}
\setlength{\tabcolsep}{1.7 em}
\begin{center}
\begin{tabular}{lcc|ccc}
\hline 
\hline
\vspace{1mm}
Type  & $\psi$ & $\delta/\pi$ & $\sin^2\theta_{23}$ & $\delta_{CP}/\pi$ & $J_{cp}$ \\
\hline 
\hline\\
\vspace*{1mm}\\
$1$& $2.9521^{+0.2087}_{-0.2043}$ & $1.764^{+0.0476}_{-0.0428}$ & $0.4585^{+0.08543}_{-0.08646}$ & $1.2308^{+0.0692}_{-0.0717},$ & $ 0.0250^{+0.0137}_{-0.0105}$   \\
\vspace*{1mm}
\vspace*{1mm}\\
$2$& $2.9521^{+0.2087}_{-0.2043}$ & $1.764^{+0.0476}_{-0.0428}$ & $0.4174^{+0.0921}_{-0.0937}$ & $1.2159^{+0.0754}_{-0.0733},$ & $ 0.0250^{+0.0137}_{-0.0105}$  \\ 
\vspace*{1mm}
\vspace*{1mm}\\
$4$& $2.9522^{+0.2087}_{-0.2201}$ & $0.7644^{+0.0476}_{-0.0427}$ & $0.4585^{+0.0855}_{-0.08641}$ & $1.2303^{+0.0717}_{-0.0713}$ & $ 0.0250^{+0.0137}_{-0.0105}$  \\ 
\vspace*{1mm}
\vspace*{1mm}\\
$4$& $2.9522^{+0.2087}_{-0.2201}$ & $0.7644^{+0.0476}_{-0.0427}$ & $0.4996^{+0.0927}_{-0.0935}$ & $1.2303^{+0.0717}_{-0.0713}$& $ 0.0250^{+0.0137}_{-0.0105}$ \\
\vspace*{1mm}\\
\hline
\end{tabular} 
\caption{\footnotesize Summary of the results corresponding to four BL models. $\psi$ and $\delta$ corresponds to the central $\pm\, 3\sigma$ range of $s_{12}^2$, $s_{13}^2$, $\lambda$ and $A$. We have taken, $s_{12}^2=[0.278,0.375]$, $s_{13}^2=[0.0177,0.0297]$, $\lambda=[0.22551-0.001,0.22551+0.001]$ and $A=[0.813-0.029,0.813-0.040]$. The other observables $s_{23}^2$, $\delta_{cp}$ (the Dirac type CP phase) and $J_{cp}$ (Jarkslog invariant parameter) are the theoretical predictions for each model. This is to be noted that the best result of the Type-4 BL model is consistent with the maximal mixing prediction.}
\label{table2}
\end{center}
\end{table*}

How to choose $\psi$ and $\delta$? In fact, this task is not too
complicated. One can choose $\psi$ and $\delta$ in such a way, that
any two of the three observable parameters, solar, reactor and
atmospheric mixing angles are consistent with the neutrino oscillation
data~\cite{Tortola:2012te,Forero:2014bxa}, while the prediction for
the remaining one will determine the tenability of the model.

First note that the determination of solar and reactor angles is
rather stable irrespective of the neutrino mass spectrum. Hence it
seems reasonable to use solar and reactor angles for the
parametrization of the two unknowns.
Hence we focus upon the predictions for $\theta_{23}$ and $J_{CP}$ (or
$\delta_{CP}$), given their current indeterminacy from global
neutrino oscillation data analysis \cite{Forero:2014bxa}. Although
consistent with maximal mixing, the possibility of $\theta_{23}$ lying
within the first octant is certainly not excluded for normal ordering
of neutrino masses. Moreover, probing for CP violation in the lepton
sector is the next challenge for neutrino oscillation
experiments. Hence in addition to the prediction for the atmospheric
angle, we use the prediction of our ansatz for $J_{CP}$ (or
$\delta_{CP}$) in order to scrutinize the viability of our ansatz, in
any of the above forms. The results are summarized in
Table\,\ref{table2}.
\begin{figure*}[h!]
\begin{center}
\includegraphics[scale=0.68]{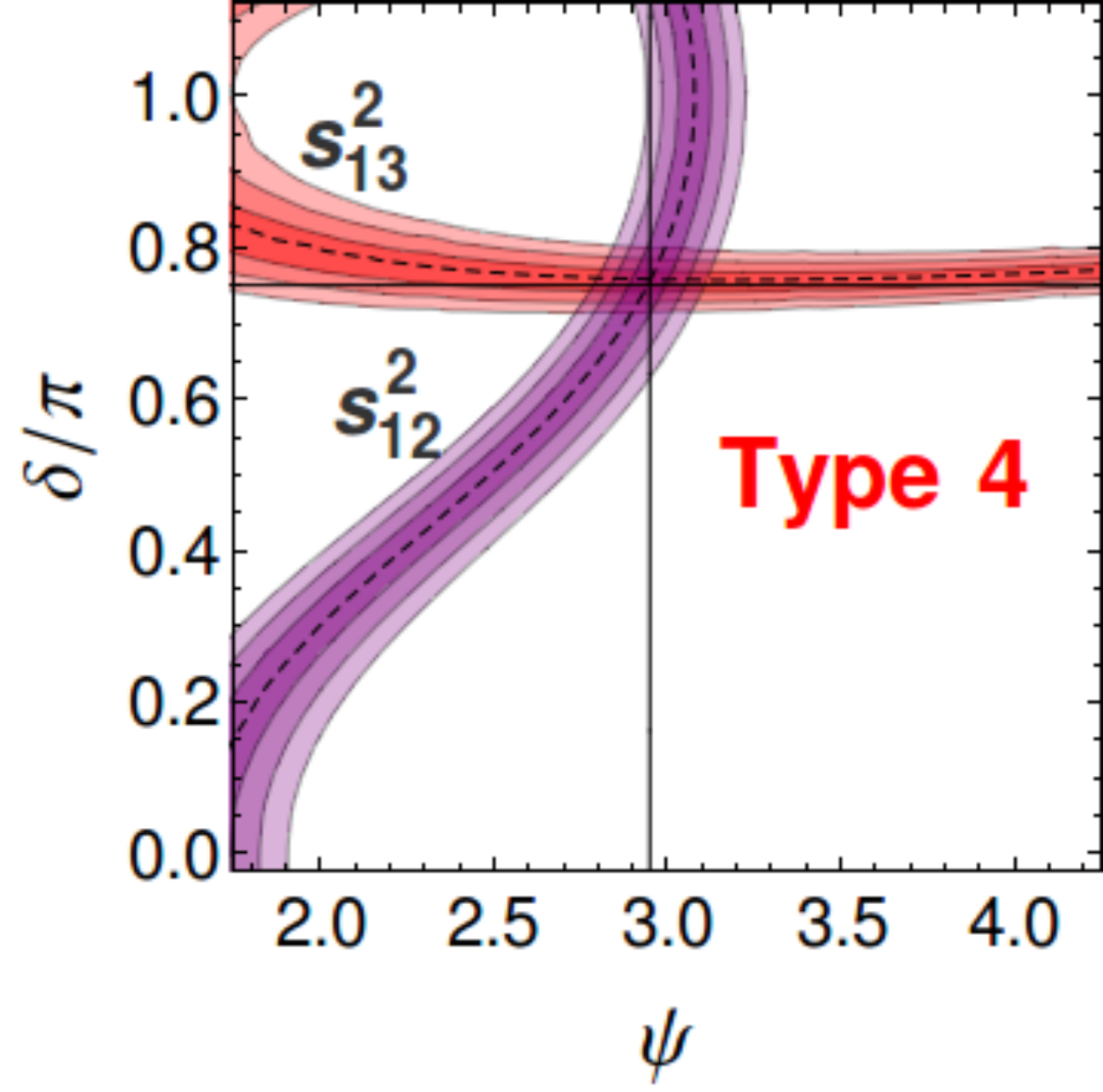}
\includegraphics[scale=0.7]{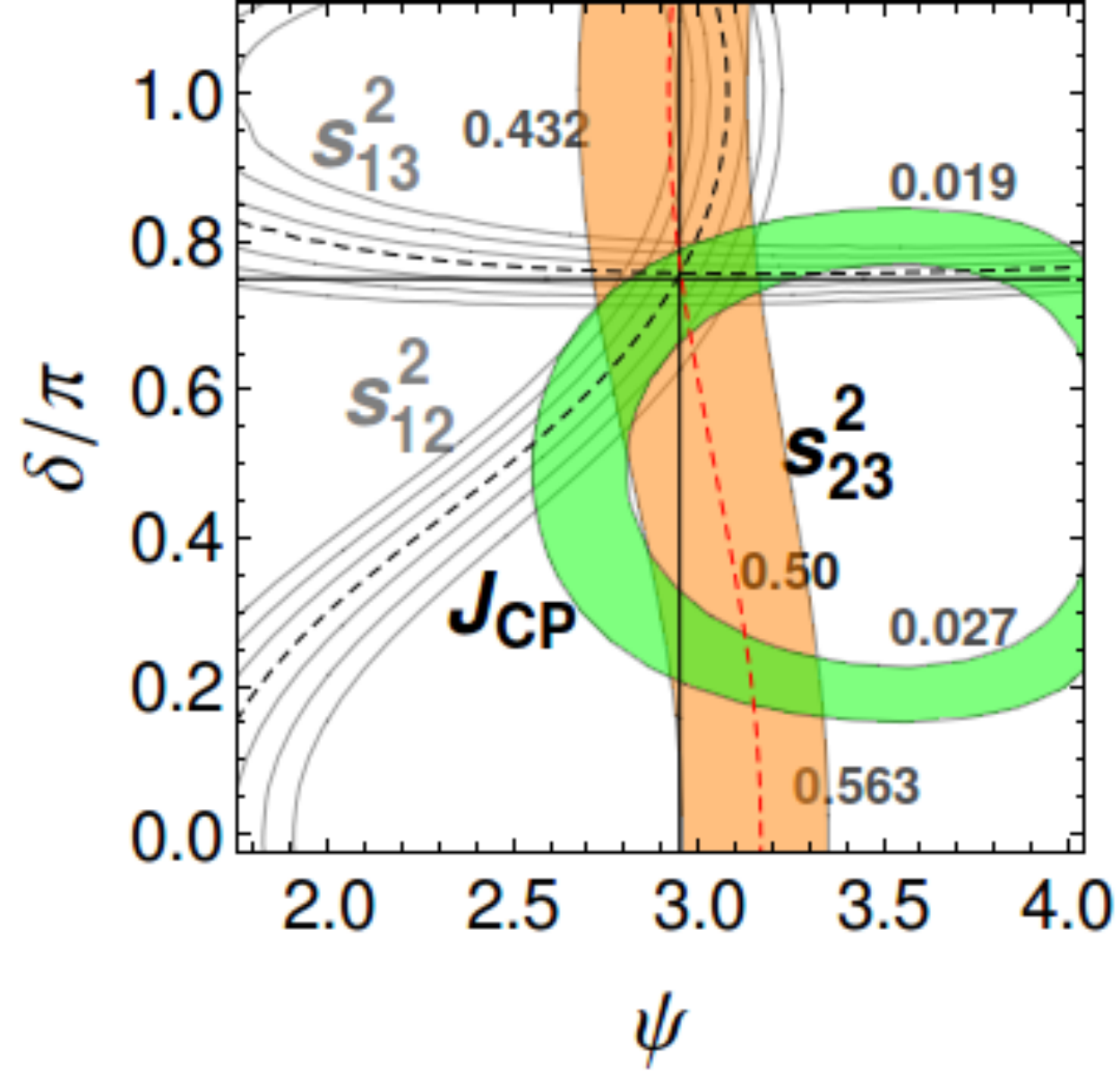}
\caption{The parametrization of $\psi$ and $\delta$, and prediction on
  $s_{23}^2$ and $J_{cp}$ are illustrated for Type-4 BL.  For all the
  cases, $\psi$ and $\delta$ are first parametrized with respect to
  best -fit, $1\sigma$, $2\sigma$ and $3\sigma$ ranges of $s_{12}^2$
  and $s_{13}^2$ which are then are used to predict $s_{23}^2$ and
  $J_{cp}$. In the above illustration we fix
    $\lambda$ and $A$ at their central values: $\lambda=0.22551$ and
    $A=0.813$ }
\label{fig4}
\end{center}
\end{figure*}
For definiteness we discuss here in more detail only the result for
the type-4 BL scheme, see Fig.\,(\ref{fig4}), similar results can be found
for the other cases in the Table.  In Fig.\,(\ref{fig4}) we plot the
free parameters $\delta$ and $\psi$. In the left panel we show the
contour plot for $s_{13}$ (horizontal band) and $s_{12}$ (vertical
band). The best fit value $s_{12}^2 \approx 0.323$ and
$s_{13}^2\approx 0.023$ \cite{Forero:2014bxa}  correspond to
choosing $\psi \approx 2.967$ and $\delta\approx 0.757\,\pi$.  We note
that, with above choice of the two parameters, $\theta_{23}$ is
consistent with maximal. The CP-invariant $J_{cp}$ is approximately
$0.02$. \\

The corresponding lepton mixing matrix corresponding to
  the Type-4 BL scheme is the following,
\begin{widetext}
\begin{align}
U_4\approx\begin{bmatrix}
-u^{*}(1+\lambda)\lbrace u(\lambda-1)+\psi \lambda^2 \rbrace c & (\psi-u^{*}c^2)\lambda +\psi\lambda^3 & \lambda-(\frac{\lambda}{2}+u^{*}\psi)\lambda^2 \\ 
\frac{c\lambda}{2}\lbrace(\lambda^2-2)(u+\psi)-2\lambda(\psi+cA\lambda)\rbrace & c^2(1-\frac{\lambda^2}{2})-\psi\lambda^2\lbrace u + (\psi + c A)\lambda\rbrace & \psi\lambda(1-\lambda^2)+(cA-u)\lambda^2 \\ 
\lbrace \psi^2 + cA(u+\psi)\lambda\rbrace\lambda^2-c^2 & -c\lbrace \psi + (\psi+cA)\lambda\rbrace\lambda & -\psi A \lambda^3 + c(1-\frac{\lambda^2}{2})
\end{bmatrix} ,
\end{align}
\end{widetext}
where $u=e^{i\delta}$ and $c=\cos\sin^{-1}(\psi\lambda)$. \\

In Table.\,\ref{table2},  we gather the results for all the four BL
schemes discussed above.\\ 

In summary we proposed a generalized fermion mixing ansatz where the
neutrino mixing is Bi-Large, while the charged lepton mixing matrix is
CKM-like.  Inspired by SO(10) and SU(5) unification, we select four
CKM-like charged lepton diagonalizing matrices, $U_{l}$'s
(Type-1,2,3,4) and discuss the phenomenological viability of
the resulting schemes.
All the four models are congruous with best-fit solar and reactor
angles, making definite predictions for the atmospheric angle and CP
phase, which may be further tested in upcoming neutrino experiments.
In particular the Type-4 BL model appears interesting in the
sense that it extends the original BL model to encompass maximal
atmospheric mixing.
Ours is a ``theory-inspired'' bottom-up approach
to the flavour problem, that highlights the role of $\theta_c$ as the
universal seed of quark and lepton mixings and incorporates the main
characteristic features of unification models.
  We have shown how this generalizes the original Bi-Large
  ansatz~\cite{Boucenna:2012xb} to make it fully realistic.
  Further investigation on the physics underlying this ansatz may
  bring new insights on both fermion mixing and unification.

  Work supported by the Spanish grants FPA2011-22975 and Multidark
  CSD2009-00064 (MINECO), and PROMETEOII/2014/084 (Generalitat
  Valenciana).  SM thanks DFG grant WI 2639/4-1.

\bibliography{merged,newrefs,refs}

\end{document}